\documentclass[12pt]{iopart}

\usepackage{color}
\usepackage{graphicx}
\usepackage{epsfig}
\usepackage{bm}

\expandafter\let\csname equation*\endcsname\relax

\expandafter\let\csname endequation*\endcsname\relax

\usepackage{amsmath}    
\usepackage{verbatim}    
\usepackage{hyperref}   

\usepackage{mathptmx}
\long\def\/*#1*/{}

\usepackage{cite}

\begin{document}

\title[]{Exploring the hadron resonance gas phase on the QCD phase diagram}

\author{Subhasis Samanta$^{1,a}$, Sandeep Chatterjee$^{2,3,b}$, Bedangadas Mohanty$^{1,4,c}$}

\address{$^1$School of Physical Sciences, National Institute of Science 
Education and Research, Bhubaneswar, HBNI, Jatni, 752050, India}
\address{$^2$ AGH University of Science and Technology, Faculty of Physics and 
Applied Computer Science, al. Mickiewicza 30, 30-059 Krakow, Poland}

\address{$^3$ Department of Physical Sciences, Indian Institute of Science Education and Research,
Berhampur, Transit Campus, Government ITI, Berhampur 760010, Odisha, India}

\address{$^4$ Experimental
Physics Department, CERN, CH-1211, Geneva 23, Switzerland}

\ead{$^a$subhasis.samant@gmail.com, $^b$Sandeep.Chatterjee@fis.agh.edu.pl, $^c$bedanga@niser.ac.in}

\vspace{10pt}

\begin{abstract}
Lattice computations of strongly interacting matter at finite temperature $T$ and baryon chemical 
potential $\mu_B$ suggest that the QCD thermodynamics deep in the hadronic 
phase can be adequately modeled by an ideal hadron resonance gas (I-HRG). However, 
it is not clear where on the $(\mu_B, T)$ plane this description breaks down, 
making it essential to account for hadronic interactions and change in the 
nature of the degrees of freedom. We have studied several thermodynamic functions within 
the I-HRG model and try to identify the region of the QCD phase diagram where it becomes 
essential to include non-ideal effects into the I-HRG model. We work with 
only those thermodynamic quantities that show a monotonic rise with $T$ and 
$\mu_B$ in I-HRG. Their high temperature limiting values where QCD becomes simply a 
Stefan-Boltzmann (SB) gas of massless quarks and gluons is known. The rise of these 
quantities in I-HRG beyond the corresponding SB limit values indicate the need to include 
interactions into I-HRG to study QCD thermodynamics. This works as a guiding principle on 
the QCD phase diagram where interacting HRG can take over from I-HRG. For $\mu_B/T\leq2$, 
$\chi^Q_2$ shoots the SB limit at the smallest $T$, while for higher values of $\mu_B/T$,
$C_{BS}=-3\chi^{BS}_{11}/\chi^S_2$ takes over. We further comment on the relative positions 
between the freezeout curve obtained by thermal fits to the measured hadron yields and the 
obtained line where I-HRG overshoots SB limit.
\end{abstract}

\section{\label{sec:Intro} Introduction}
Substantial theoretical and experimental efforts world wide have been
devoted to the investigation of strongly interacting matter under
extreme temperature $T$ and baryon chemical potential $\mu_B$. Lattice quantum
chromodynamics (LQCD) ~\cite{Aoki:2006we,Borsanyi:2011sw,
  Gupta:2011wh,  Bazavov:2012jq,Bellwied:2013cta, Borsanyi:2013bia,
  Bazavov:2014pvz, Bellwied:2015lba, Ding:2015ona} calculation suggests a smooth crossover transition ~\cite{Aoki:2006we} from 
hadronic to a quark-gluon plasma (QGP) phase at zero $\mu_{B}$ and finite 
$T$~\cite{Gupta:2011wh}. Depending on the choice of order parameter the transition
occurs approximately between the temperature 145 MeV to 175 MeV at
zero $\mu_{B}$. For example, LQCD calculation with chiral condensate
as order parameter gives 
$T_c \sim 154$ MeV \cite{Bazavov:2011nk}. However if one chooses
strange quark number susceptibility then the $T_c \sim$ 170 MeV
\cite{Borsanyi:2010bp}. Various QCD based model calculations at high
baryon density and low temperature suggest the existence of a first-order phase
transition ~\cite{Asakawa:1989bq}. Computing LQCD calculations at high 
$\mu_{B}$ is numerically challenging. 
Hence at high $\mu_{B}$, there are large
uncertainties in calculating the transition line across the $T$ versus
$\mu_{B}$ phase diagram of QCD 
\cite{Endrodi:2011gv,Kaczmarek:2011zz,Bonati:2014rfa,Cea:2014xva,
Bonati:2015bha}. Although at small $\mu_{B}$, precise computation
of transition line has been carried out recently~\cite{Bazavov:2018mes}.
Several experimental programs have
been devoted to find this transition line in the phase diagram of
QCD. At present, QCD system at high $T$ and small $\mu_B$ are being investigated using 
ultra-relativistic heavy ion collisions at the Large Hadron Collider (LHC), CERN
and Relativistic Heavy Ion Collider (RHIC), Brookhaven National Laboratory (BNL).
The Beam Energy Scan (BES) program of RHIC \cite{Abelev:2009bw} is currently
investigating the matter at a wide range of $\mu_{B}$ : 20 to 400 MeV \cite{Adamczyk:2013dal}.
The HADES experiment at GSI, Darmstadt is also investigating a medium with very large baryon chemical potential
\cite{Agakishiev:2015bwu}. In future, the Compressed Baryonic Matter (CBM) experiment \cite{Ablyazimov:2017guv}
at the Facility for Antiproton and Ion Research (FAIR) at GSI and the
Nuclotron-based Ion Collider Facility (NICA) \cite{Kekelidze:2017ual} at JINR, 
Dubna will also study nuclear matter at large baryon chemical potential. 

Hadron resonance gas (HRG) model 
 ~\cite{Hagedorn:1980kb, Rischke:1991ke,Cleymans:1992jz, BraunMunzinger:1994xr, Cleymans:1996cd, Yen:1997rv,
 BraunMunzinger:1999qy, Cleymans:1999st, BraunMunzinger:2001ip, BraunMunzinger:2003zd, Karsch:2003zq, Tawfik:2004sw,
Becattini:2005xt,  Andronic:2005yp, Andronic:2008gu,Begun:2012rf, Andronic:2012ut,
Tiwari:2011km, Fu:2013gga, Tawfik:2013eua, Garg:2013ata, Bhattacharyya:2013oya,
Bhattacharyya:2015zka,Chatterjee:2013yga,Chatterjee:2014ysa,Chatterjee:2014lfa,Becattini:2012xb,Bugaev:2013sfa,
Petran:2013lja, Vovchenko:2014pka, Kadam:2015xsa, Kadam:2015fza, Albright:2014gva, Albright:2015uua,
Bhattacharyya:2015pra, Kapusta:2016kpq, Begun:2016cva,Adak:2016jtk, Xu:2016skm,Fu:2016baf,
Vovchenko:2015xja, Vovchenko:2015vxa, Vovchenko:2015pya, Broniowski:2015oha, Vovchenko:2015idt,
Redlich:2016dpb, Vovchenko:2016rkn, Alba:2016fku, Samanta:2017kmg, Samanta:2017ohm, Sarkar:2017ijd, 
Bhattacharyya:2017gwt, Chatterjee:2017yhp, Alba:2016hwx, Samanta:2017yhh}
is a statistical thermal model which is used to study the strongly interacting matter at finite temperature 
and chemical potential.
The HRG model is successful in describing the zero chemical potential LQCD data of bulk
properties of the QCD matter upto  moderate temperatures $T \sim 150$ MeV
~\cite{Borsanyi:2011sw, Bazavov:2012jq, Bazavov:2014pvz,Bellwied:2013cta, Bellwied:2015lba}.
This model is also successful in describing the hadron yields,
created in central heavy ion collisions at different center of mass
energies ($\sqrt{s_{NN}}$)
~\cite{BraunMunzinger:1994xr, Cleymans:1996cd, Cleymans:1999st, 
Andronic:2005yp,Adamczyk:2017iwn}, at chemical freeze-out which is the stage in the evolution of the thermal
system when inelastic collisions among the hadrons cease and the hadronic yields become fixed.
The values of $T$ and $\mu_{B}$ extracted using the HRG model at large
$\sqrt{s_{NN}}$ is very close to $T_c$ obtained from LQCD calculations
at zero $\mu_{B}$. This raises the interesting question - Is the
chemical freeze-out line same as the hadronization or quark-hadron
transition line ?  In order to address this one needs LQCD calculations (which are
difficult to compute numerically at large $\mu_{B}$), precise
experimental signatures related to transition line at several
$\sqrt{s_{NN}}$ and study of chemical freeze-out dynamics using HRG
model in its different variants.  In this work we ask a slightly different yet
related question - What is the upper limit in the $T$ versus $\mu_{B}$
phase diagram upto which the QCD thermodynamics can be effectively modeled 
by an ideal HRG (I-HRG) model? We choose only 
those thermodynamic quantities (TQs) that show a monotonic behavior between the 
hadronic and quark gluon plasma phase with limiting values corresponding to 
that computed for the Stefan-Boltzmann (SB) gas of massless quarks and gluons. 
For a given $\mu_B$, we find the $T$ where the TQ computed in I-HRG exceeds the 
corresponding SB limit. This process is repeated for several TQs. 
A trace of the line joining the lowest $T$ values at each $\mu_{B}$ from 
all the observables studied provides the upper limit for the allowed 
region of the I-HRG phase on the QCD phase diagram. 
Here we want to mention that for some TQs upper limit is loose 
which can be therefore improved through first-principle LQCD computations.

The paper is organized as follows. In Sec \ref{sec:Model} we briefly discuss
the hadron resonance gas model and the calculation related to the SB limit. In 
Sec. \ref{sec:Results} we discuss our results and finally in Sec. \ref{sec:Summary} 
we summarize our findings for this work.

\section{\label{sec:Model} Ideal Hadron Resonance Gas Model}

The system of thermal fireball consists of all the hadrons and resonances given 
in the particle data book \cite{Patrignani:2016xqp}.
In this model hadrons and resonances are non-interacting point like particles.
The partition function is the basic quantity from which one can calculate various TQs of 
the thermal system. The logarithm of the partition function of an ideal
hadron resonance gas in the grand canonical ensemble can be written as \cite{Andronic:2012ut}
\begin {equation}
 \ln Z^{H}=\sum_i \ln Z_i^{H},
\end{equation}
where the sum is over all the hadrons and resonances and $H$ refers to the hadronic phase.
For the hadron or resonance species $i$,
\begin{equation}
 \ln Z_i^{H}=\pm \frac{Vg_i}{2\pi^2}\int_0^\infty p^2\,dp \ln[1\pm\exp(-(E_i-\mu_i)/T)],
\end{equation}
where $V$ is the volume of the thermal system, 
$g_i$ is the degeneracy, $E_i = \sqrt{p^2 + m_i^2}$
is the single particle energy, $m_i$ is the mass of the particle
and $\mu_i=B_i\mu_B+S_i\mu_S+Q_i\mu_Q$ is the
chemical potential. In the last expression, $B_i,S_i,Q_i$ are respectively
the baryon number, strangeness and electric charge of the particle, $\mu^,s$ are 
the corresponding chemical potentials.
The upper and lower sign of $\pm$ corresponds
to fermions and bosons, respectively.
Once we know the partition function or the pressure of the system
we can calculate other TQs.
The pressure $P^{H}$, the number density $n^{H}$, the energy density 
$\varepsilon^{H}$, and the entropy density $s^{H}$ of the system
can be calculated using the standard definitions \cite{Andronic:2012ut},


\begin{equation}
 P^{H}=\sum_i \frac{T}{V}\ln Z_i^{H}
  =\sum_i (\pm)\frac{g_iT}{2\pi^2}\int_0^\infty p^2\,dp \ln[1\pm\exp(-(E_i-\mu_i)/T)],
\end{equation}
 
%

\begin{equation}\label{eq:n}
 n^{H}= \sum_i \frac{T}{V} \left(\frac{\partial \ln Z_i^{H}}{\partial\mu_i}\right)_{V,T}
 =\sum_i \frac{g_i}{2\pi^2}\int_0^\infty\frac{p^2\,dp}{\exp[(E_i-\mu_i)/T]\pm1},
\end{equation}

\begin{equation}
 \varepsilon^{H}=\sum_i \frac{E_i^{H}}{V}=- \sum_i \frac{1}{V} \left(\frac{\partial \ln Z_i^{H}}{\partial\frac{1}{T}}\right)_{\frac{\mu}{T}}
=\sum_i \frac{g_i}{2\pi^2}\int_0^\infty\frac{p^2\,dp}{\exp[(E_i-\mu_i)/T]\pm1}E_i,
\end{equation}



 \begin{align}\label{eq:s}
  \begin{split}
   s^{H}&=\sum_i \frac{S_i^{H}}{V}=\frac{1}{V}\left(\frac{\partial\left({T \ln Z_i^{H}}\right)}{\partial T}\right)_{V,\mu}\\
  &=\sum_i (\pm) \frac{g_i}{2\pi^2}\int_0^\infty p^2\,dp \left[ \ln\left(1\pm\exp(-\frac{(E_i-\mu_i)}{T})\right)\right.
  \left.\pm\frac{(E_i-\mu_i)}{T(\exp((E_i-\mu_i)/T)\pm1)}\right].
  \end{split}
 \end{align}
 

Fluctuations and correlations of conserved charges of baryon number, electric charge, 
strangeness and others are considered as a standard probe
to study the phase transition.
Derivatives of the grand canonical partition function ($Z$) with respect to the chemical potential 
define susceptibilities which experimentally become available through event-by-event analysis 
of fluctuations of conserved charges \cite{Adamczyk:2013dal,Adamczyk:2014fia,Adare:2015aqk}.
For example, second order fluctuations of the conserved charges and their correlations 
in a thermalized medium can be calculated as 
\begin{equation}
 \chi^2_x=\frac{1}{V T^3}\frac{\partial^2 {(ln Z)}}{\partial {(\frac{\mu_x}{T})}^2},
\end{equation}

\begin{equation}
 \chi^{11}_{xy}=\frac{1}{V T^3}\frac{\partial^{2} {(ln Z)}}{\partial {(\frac{\mu_x}{T})} \partial {(\frac{\mu_y}{T})}},
\end{equation}
where $x, y = B$ (baryon),
$S$ (strangeness) and $Q$ (electric charge).

In order to compare theoretical computation with that obtained in experiments, suitable ratios are formed to cancel 
the system volume. In this work, apart from the various TQs, we also work with the following ratio 
that is expected to show a monotonic variation
\begin{equation}
C_{BS} = -3 \left(\chi^{11}_{BS}\right)^{}/\left(\chi^2_S\right)^{}.
\end{equation}

\section{Ideal gas of quarks and gluons}
The pressure of a massless quark gluon gas of three flavor QCD [Stefan-Boltzmann (SB) gas]
can be written as \cite{Muller:1983ed}
\begin{equation}\label{SB_P}
 \frac{P^{SB}}{T^4} = \frac{8 \pi^2}{45} + \sum_{f=u,d,s} \frac{7 \pi^2}{60} 
 + \frac{1}{2} \left(\frac{\mu_f}{T}\right)^2 + \frac{1}{4 \pi^2} \left(\frac{\mu_f}{T}\right)^4,
\end{equation}
where the two terms give the contributions of the gluon and the quark sector respectively,
$\mu_f$ is the quark chemical potential. 
The quark chemical potentials of $u, d$ and $s$ quark can be expressed in terms of
chemical potentials for baryon number ($\mu_B$), strangeness ($\mu_S$) and
electric charge ($\mu_Q$) as 

\begin{equation}
 \mu_u = \frac{1}{3} \mu_B + \frac{2}{3} \mu_Q
\end{equation}

\begin{equation}
 \mu_d = \frac{1}{3} \mu_B - \frac{1}{3} \mu_Q,
\end{equation}

\begin{equation}
 \mu_s = \frac{1}{3} \mu_B - \frac{1}{3} \mu_Q - \mu_S.
\end{equation}


After knowing the $P^{SB}/T^4 (T, \mu_B, \mu_S, \mu_Q)$ or the corresponding partition function
one can calculate other TQs as well as the fluctuations and correlations of 
different conserved charges of a free quark gluon gas at any 
temperature and chemical potential using the definitions mentioned in the previous section.
The energy density and the entropy density of the massless quark gluon gas is given by
\begin{equation}\label{SB_E}
 \frac{\varepsilon^{SB}}{T^4} = 3 \frac{P^{SB}}{T^4},
\end{equation}

\begin{equation}\label{SB_s}
 \frac{s^{SB}}{T^3} = \frac{19 \pi^2}{9} + \sum_{f=u,d,s} \left(\frac{\mu_f}{T}\right)^2.
\end{equation}

Second order fluctuations of baryon, strangeness and electric charge at the mass less limit
of the quark gluon gas can be written as
\begin{equation}
 \left(\chi^2_B\right)^{SB} = \frac{1}{3} + \frac{1}{3\pi^2 T^2} \sum_{f=u,d,s} \mu_f^2,
\end{equation}

\begin{equation}\label{eq:chi_s2_sb}
 \left(\chi^2_S\right)^{SB} = 1 + \frac{3}{\pi^2 T^2} \mu_s^2,
\end{equation}

\begin{equation}
 \left(\chi^2_Q\right)^{SB} =  \frac{2}{3} + \frac{1}{3\pi^2 T^2} \left( 4\mu_u^2 + \mu_d^2 + \mu_s^2\right).
\end{equation}

Similarly baryon-strangeness and charge-strangeness correlations are given by

\begin{equation}\label{eq:chi_bs_sb}
 \left(\chi^{11}_{BS}\right)^{SB} = -\left(\chi^{11}_{QS}\right)^{SB} = -\frac{1}{3} - \frac{1}{\pi^2 T^2} \mu_s^2.
\end{equation}

It can be seen from Eqs. \ref{eq:chi_s2_sb} and \ref{eq:chi_bs_sb}
that $C_{BS}^{SB} = -3 \left(\chi^{11}_{BS}\right)^{SB}/\left(\chi^2_S\right)^{SB}$ is
always unity for the quark gluon gas.

\begin{figure}[]
\centering
\includegraphics[width=0.45\textwidth]{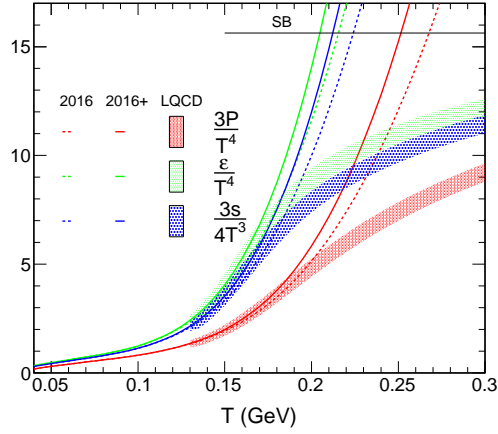}
\vspace{0.5cm}
\caption{Variation of $3 P/T^4$, $\varepsilon/T^4$ and $3 s/4T^3$ with temperature at
zero $\mu_B$. Dotted and solid lines correspond to the calculations in HRG model with 
hadron spectra from PDG 2016 and 2016+. Red, green and blue colors are used for pressure, 
energy density and entropy density respectively. The colored shaded bands show the 
continuum estimate of the lattice QCD \cite{Bazavov:2014pvz}. Horizontal line at $19\pi^2/12$ 
is the corresponding SB limit.}
\label{fig:eos_mu0}
\end{figure}

\section{\label{sec:Results}Results}

In this work, we aim to explore the I-HRG phase and try to estimate the region on 
the $\mu_B-T$ plane where it is a good approximation to QCD. For our analysis, we 
investigate the pressure, energy density, entropy density and several second order fluctuations 
and correlations of conserved charges that increase monotonically in the I-HRG model. 
On the other hand, in real QCD these TQs are expected to attain 
limiting values corresponding to the SB limit at sufficiently large $T$ and $\mu_B$. Thus, 
the monotonic rise in I-HRG of the TQs beyond their SB-limit values indicate possible 
pathology in the model rendering it unsuitable to describe QCD thermodynamics. 

It possibly signals 
the requirement to go beyond the I-HRG model in describing the QCD 
thermodynamics like incorporating hadron interactions~\cite{Huovinen:2017ogf, Vovchenko:2017zpj}, in-medium 
modification of hadron properties~\cite{Aarts:2018glk}, accounting 
for hadron melting~\cite{Jakovac:2013iua} and change in the relevant degrees of freedom etc. We identify the 
$(\mu_B,T)$ curve corresponding to each TQ beyond which the I-HRG value exceeds those of the 
corresponding SB-limit values. This provides an estimate of the region of applicability of the 
I-HRG model to understand QCD thermodynamics.

\subsection{\label{sec:Result1} $\mu_{B,Q,S} = 0$}

Let us first discuss the picture for zero chemical potential. In Fig. \ref{fig:eos_mu0} we 
show the variation of $3 P/T^4$, $\varepsilon/T^4$ and $3 s/4T^3$ with $T$. The TQs are normalized 
suitably so that they have the same SB limit of $19\pi^2/12$. We have done the calculations in I-HRG 
with two different input hadronic spectra. For the first set, we have considered all the confirmed 
hadrons and resonances that consist only up, down and strange flavor valence quarks listed in the 
PDG 2016 Review \cite{Patrignani:2016xqp}. This list includes all the confirmed mesons listed in the 
Meson Summary Table \cite{Patrignani:2016xqp} and all baryons in the Baryon Summary Table \cite{Patrignani:2016xqp} 
with three- or four-star status. We refer to this set as PDG 2016. The dotted lines show the result 
of I-HRG using hadronic spectrum PDG 2016. The red, green and blue colors are used for pressure, energy 
density and entropy density respectively. We have done our analysis for another set of hadronic spectrum. 
This set includes all the resonances from the previous set i.e., PDG 2016 as well as the other unmarked 
mesons from the Meson Summary Table and baryons from the Baryon Summary Table with one- or two-
star status which are not confirmed yet \cite{Patrignani:2016xqp}. This set is referred to as
PDG 2016+ and is plotted in solid lines in Fig. \ref{fig:eos_mu0}. The colored shaded bands show the 
continuum estimate from lattice QCD \cite{Bazavov:2014pvz}. The hadronic spectrum PDG 2016+ provides 
a satisfactory description in the hadronic phase of continuum LQCD data of most of the
TQs which is already known from the previous work \cite{Alba:2017mqu}. Additional resonances in this 
list will allow us to study the systematic uncertainties for our analysis. It can be seen from the 
Fig. \ref{fig:eos_mu0} that pressure, energy density and entropy density calculated in HRG model using 
hadronic spectrum PDG 2016 cross the SB limit at $T = 268, 216$ and 224 MeV respectively. For hadronic 
spectrum PDG 2016+, the corresponding temperatures are $T = 252, 204$ and 212 MeV respectively. So for PDG 2016+
all the TQs reach the SB limit at relatively lower temperature compared to that of PDG 2016.
This trend of lowering of temperature due to the systematics of the hadron spectrum (mainly addition of new resonances)
is also observed in the chemical freeze-out temperature \cite{Chatterjee:2017yhp}. We also 
note that the crossing $T$ for $s$ and $\varepsilon$ which are first derivatives of $\ln Z$ are lower 
compared to that of $P$. This trend follows even for other TQs as well- higher the derivative of $\ln Z$, 
lower is the crossing $T$.

\begin{figure}[]
\centering
\includegraphics[width=0.45\textwidth]{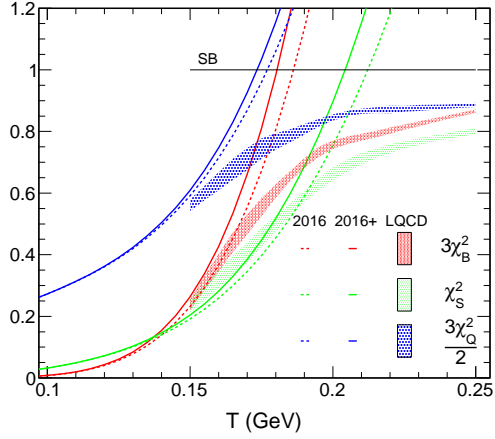}
\vspace{0.5cm}
\caption{
The variation of $3 \chi^2_B$, $\chi^2_S$ and $3 \chi^2_Q/2$
with the temperature at zero $\mu_B$. Dotted and solid lines 
correspond to the calculations in I-HRG model with hadron spectra 
PDG 2016 and 2016+. Red, green and blue colors are used for second order fluctuations of
baryon, strangeness and electric charge respectively.
The colored shaded bands show the continuum estimate of the lattice QCD \cite{Bazavov:2012jq}.
Horizontal line at $1$ corresponds to the massless limit (SB)
of $\chi^2_S$ of the three flavored quark gluon gas.
}
\label{fig:fluc_mu0}
\end{figure}

Figure \ref{fig:fluc_mu0} shows the variation of $3 \chi^2_B$, $\chi^2_S$ and $3 \chi^2_Q/2$
with the temperature at zero chemical potential.
The dotted and solid lines correspond to the calculations in I-HRG model with hadron spectra 
PDG 2016 and 2016+ respectively. 
The red, green and blue color are used for second order fluctuations of
baryon, strangeness and electric charge respectively.
The colored shaded bands show the continuum extrapolated LQCD data \cite{Bazavov:2012jq}.
For the massless three flavored quark and gluon gas values
of $\chi^2_B, \chi^2_S$ and $\chi^2_Q$ are 1/3, 1 and 2/3 respectively.
The horizontal line at $1$ in Fig. \ref{fig:fluc_mu0} 
corresponds to the SB limit of $\chi^2_S$. In this figure $\chi^2_B$ and $\chi^2_Q$ are normalized
properly so that their values at SB limit also become unity.
It can be seen from this figure
that for the hadronic spectrum PDG 2016, fluctuations of baryon, strangeness and 
charge reach the corresponding SB value at $T = 186, 212$ and 176 MeV respectively.
For the hadronic spectrum PDG 2016+, the corresponding temperatures are $T = 180, 200$ and 174 MeV 
respectively. At small $T$, the order of the masses of the lightest hadron in each charge sector 
decides the order of the magnitudes of the susceptibilities. However, with $T\sim150$ MeV, the 
stronger rise in the baryonic spectrum as compared to the meson spectrum results in $\chi^2_B$ taking 
over. Finally, both $\chi^2_Q$ and $\chi^2_B$ have similar crossing $T$ while $\chi^2_S$ has a much 
higher crossing $T$.

\begin{figure}[]
\centering
\includegraphics[width=0.45\textwidth]{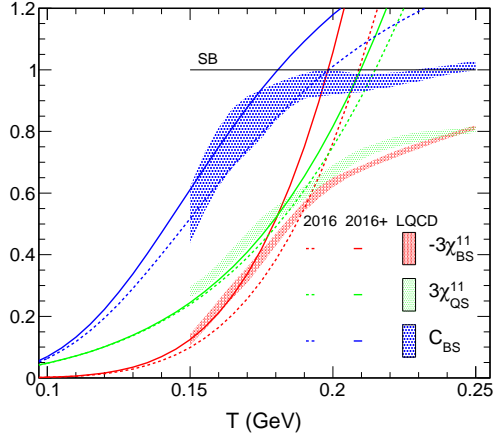}
\vspace{0.5cm}
\caption{The variation of $-3 \chi^{11}_{BS}$, $3 \chi^{11}_{QS}$ and $C_{BS}$
with the temperature at zero chemical potential.
Dotted and solid lines correspond to the calculations in HRG model with hadron spectra 
PDG 2016 and 2016+ respectively.
Red, green and blue colors are used for second order fluctuations of
baryon, strangeness and electric charge respectively.
The colored shaded bands show the continuum estimate of the lattice QCD \cite{Bazavov:2012jq}.
Horizontal line at $1$ corresponds to the massless limit (SB)
of $C_{BS}$ of the three flavored quark gluon gas.}
\label{fig:corr_mu0}
\end{figure}

Figure \ref{fig:corr_mu0} shows temperature dependence of $-3 \chi^{11}_{BS}$, $3 \chi^{11}_{QS}$ 
and $C_{BS}$ at zero chemical potential.
The results using hadronic spectrum  
PDG 2016 (2016+) are shown by the dotted (solid) lines.
Red, green and blue colors are used for $-3 \chi^{11}_{BS}$, $3 \chi^{11}_{QS}$ 
and $C_{BS}$ respectively. The continuum extracted LQCD data \cite{Bazavov:2012jq}
are shown by the colored shaded bands.
The horizontal line at $1$ corresponds to the SB limit
of all the observables shown in this figure. $\chi^{11}_{BS}$, $3 \chi^{11}_{QS}$ 
and  $C_{BS}$ calculated in HRG model using hadronic spectrum PDG 2016 cross the 
SB limit at $T = 208, 214$ and 198 MeV respectively. For the hadronic 
spectrum PDG 2016+ corresponding temperatures are $T = 186, 210$ and 180 MeV respectively. 
The influence of additional Hagedorn type resonances on $C_{BS}$ and its consequences 
on the applicability of I-HRG was discussed in Ref.~\cite{Chatterjee:2009km}.

\begin{figure}[]
\centering
\includegraphics[width=0.48\textwidth]{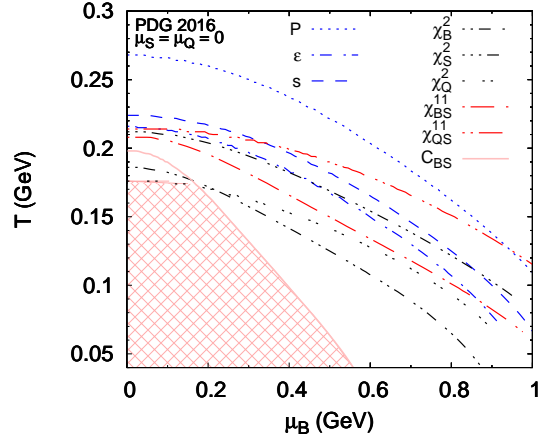}\label{fig:phase2}
\vspace{0.5cm}
\caption{Hadronic phase boundaries in the $T, \mu_B$ phase diagram at $\mu_S = \mu_Q =0$.
It is assumed that the hadronic phase ends when a particular thermodynamic quantity calculated
in the HRG model crosses the corresponding values of the mass less three flavor QGP.
That means hadronic phase ends when $O^H = O^{SB}$, where $O$ can be any observable.
Different lines correspond to the phase boundaries using different TQs.
Calculations are done with hadron spectra from PDG 2016.
The shaded region of the phase diagram is the common hadronic phase for all the observables.
}
\label{fig:phase_wo_con}
\end{figure}

\subsection{\label{sec:Result2} $\mu_B\neq0$; $\mu_S = \mu_Q =0$}

\begin{figure}[]
\centering
\includegraphics[width=0.48\textwidth]{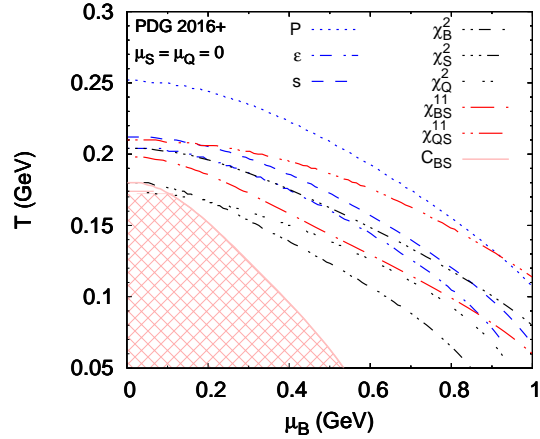}\label{fig:phase1}
\vspace{0.5cm}
\caption{Same as Fig. \ref{fig:phase_wo_con} but for a different set of hadronic spectra 
which we refer as PDG 2016+.}
\label{fig:phase_wo_con_ex}
\end{figure}

\begin{figure}[]
\centering
\includegraphics[width=0.48\textwidth]{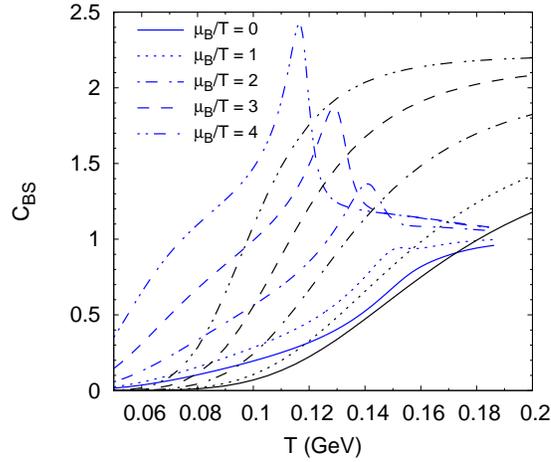}
\vspace{0.5cm}
\caption{Variation of $C_{BS}$ as function of $T$ at different $\mu_B$. Blue 
and black lines are for PQM~\cite{Chatterjee:2011jd,Chatterjee:2012np,Chatterjee:2015oka} 
and I-HRG with PDG 2016+ respectively.}
\label{fig:chi2B_T_scale}
\end{figure}

\begin{figure*}[]
\centering
\includegraphics[width=0.48\textwidth]{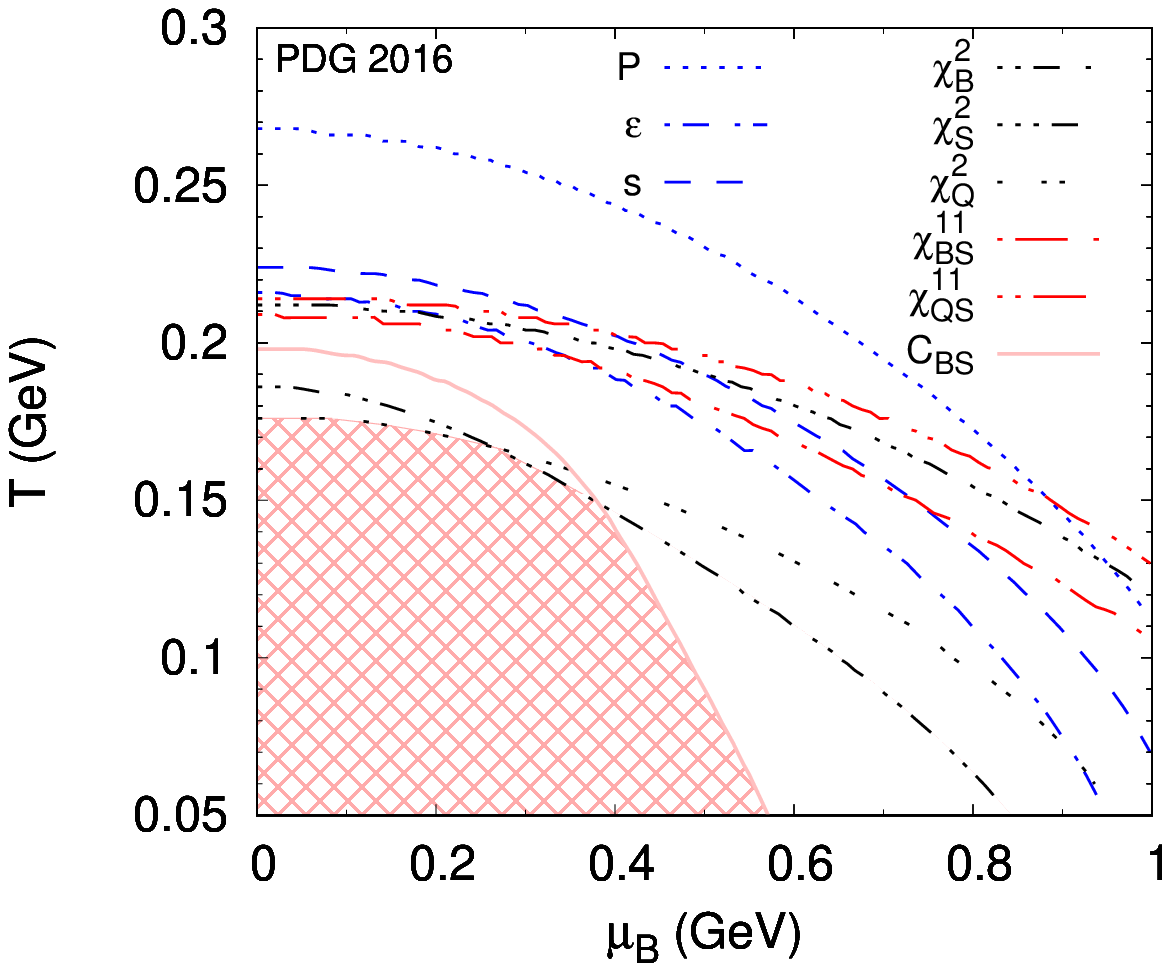}
\includegraphics[width=0.48\textwidth]{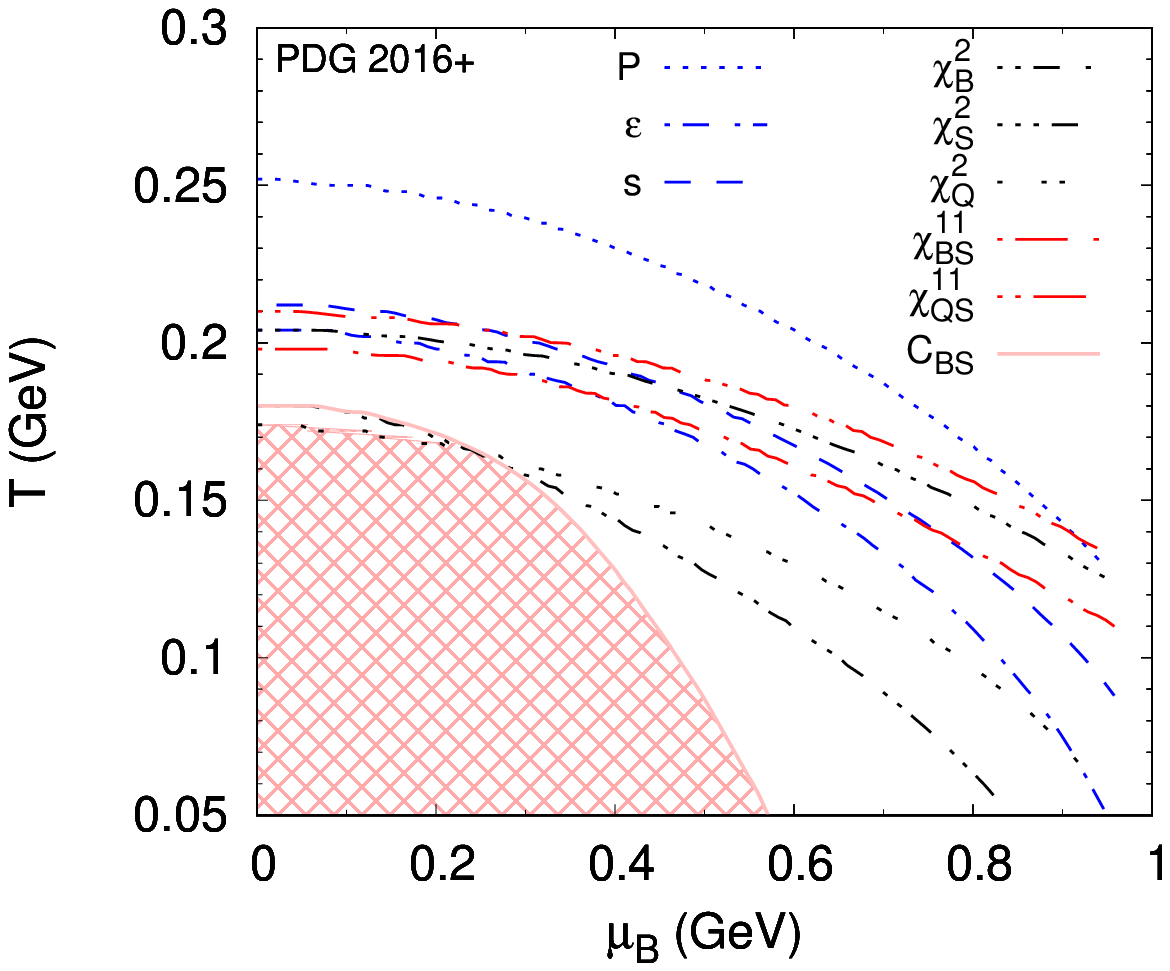}
\vspace{0.5cm}
\caption{Same as Fig. \ref{fig:phase_wo_con} or \ref{fig:phase_wo_con_ex} but for heavy-ion scenario. In this case global
charge conservations are applied to get $\mu_S$ and $\mu_Q$ at a fixed $T, \mu_B$ point. 
Left and right panels show the results using the hadronic spectrum PDG 2016 and PDG 2016+ respectively.
}
\label{fig:phase_with_con}
\end{figure*}

We now extend our analysis on the $(\mu_B, T)$ plane. In Figs.~\ref{fig:phase_wo_con} and 
\ref{fig:phase_wo_con_ex} we show the boundaries beyond which I-HRG exceed the corresponding SB limit 
values. While the results in Fig.~\ref{fig:phase_wo_con} are computed with 
hadron spectra PDG 2016, those in Fig.~\ref{fig:phase_wo_con_ex} have been done with hadron 
spectra PDG 2016+. Each TQ provides a separate boundary beyond which the I-HRG exceeds the 
corresponding SB limit. For example, along the blue dotted line 
pressure in the I-HRG phase is equal to the SB limit (i.e., $P^H = P^{SB}$). We have drawn 
similar lines using other TQs like $\varepsilon$, $s$, $\chi^2_B$, $\chi^2_S$, $\chi^2_Q$, 
$\chi^{11}_{BS}$, $\chi^{11}_{QS}$ and $C_{BS}$ obeying $O^H = O^{SB}$ where $O$ is any of 
the TQs. The shaded region is obtained by tracing out the lowest $T$ value for a given 
$\mu_{B}$ for all the TQs studied to provide the upper bound on the QCD phase diagram for the 
I-HRG model wherein the model estimate does not exceed the corresponding SB limit. With hadron 
spectra 2016+, TQs rise faster with $T$ and $\mu_B$ as compared to the case with hadron spectra 
2016 resulting in a tighter bound for the former as already seen in Figs.~\ref{fig:eos_mu0}, 
\ref{fig:fluc_mu0} and \ref{fig:corr_mu0} for $\mu_B=0$. As already noted earlier, the broad 
trend seems to be that higher the derivative of $\ln Z$ considered, the SB limit is attained 
for lower $(\mu_B, T)$ values. The curvatures of the boundaries obtained for the different TQs 
are similar except for $C_{BS}$ which has a much larger curvature than the rest resulting in the 
tightest bound on the region on the QCD phase diagram where the I-HRG model does not exceed the 
corresponding SB value.

\begin{figure}[]
\centering
\includegraphics[width=0.4\textwidth]{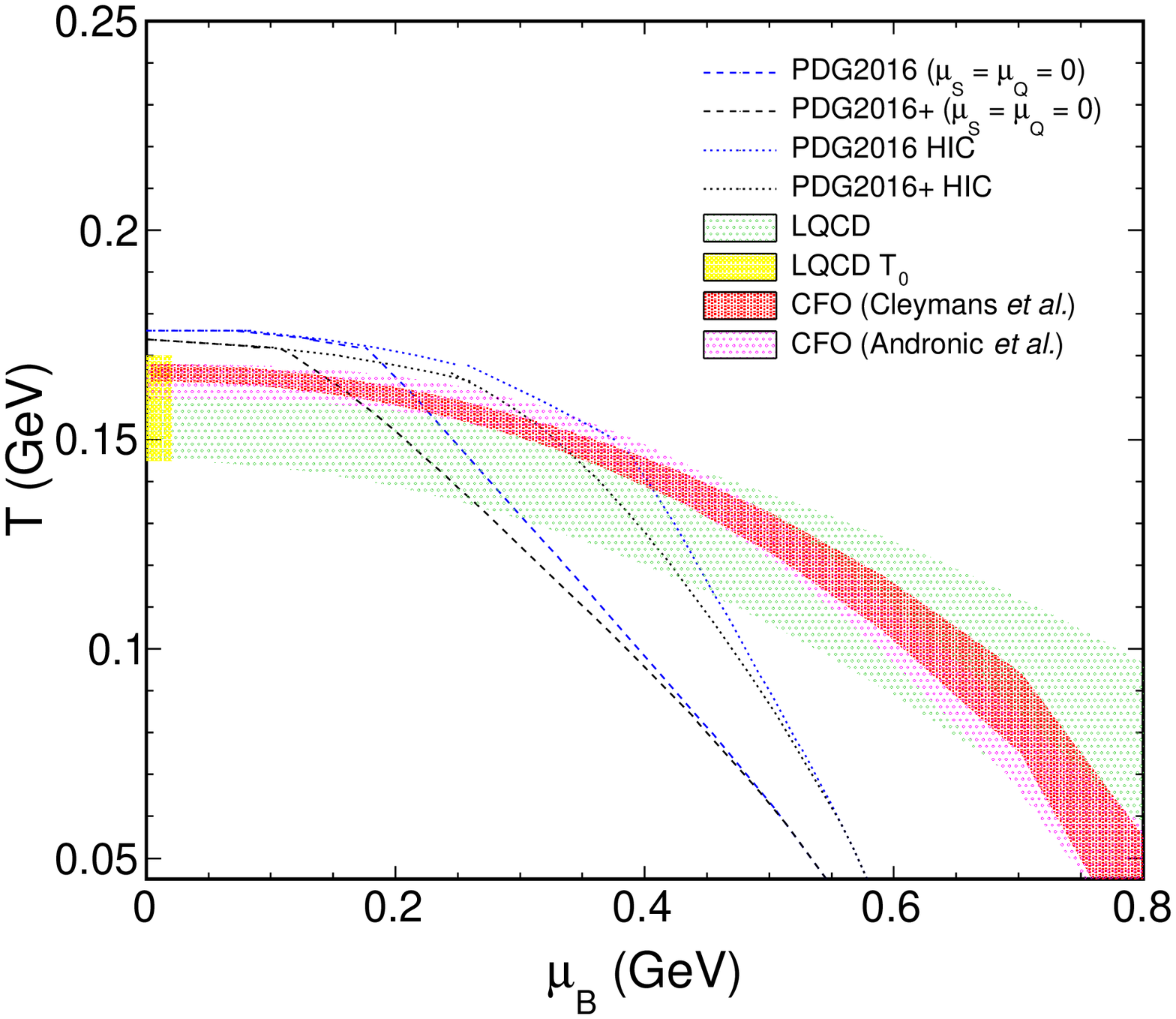}
\vspace{0.5cm}
\caption{Comparison of all hadronic phase boundaries calculated in the present work
with the lattice QCD calculation.
These are the boundaries of the shaded regions or the common hadronic phase of
 previous plots.
Blue and black dotted lines correspond to the hadronic phase boundaries at $\mu_S = \mu_Q =0$
using hadronic spectra PDG 2016 and PDG 2016+ respectively. Blue and black solid lines show
similar phase boundaries but for the heavy ion collision scenario where $\mu_S$ and $\mu_Q$
have been calculated by applying charge conservations as
mentioned in Eqs. \ref{eq:ns}- \ref{eq:nbq}.
The green error band shows the lattice transition line \cite{Cea:2014xva}.
The yellow box shows the uncertainty in 
LQCD transition temperature (145 - 170 MeV) at $\mu_B =0$.
Chemical freeze-out parameters of Refs \cite{Cleymans:2005xv} and \cite{Andronic:2008gu}
are shown by the red and pink colored shaded bands respectively.}
\label{fig:phase_com}
\end{figure}

To understand the behavior of $C_{BS}$ at large $\mu_B$ we have used a QCD like model
called the Polyakov Quark Meson (PQM) model~\cite{Schaefer:2007pw}. In Fig. \ref{fig:chi2B_T_scale} we have 
shown the variation of $C_{BS}$ as function of $T$ at different $\mu_B/T$. Blue and black 
lines are used for PQM and I-HRG respectively with PDG 2016+ hadron spectrum. We have used $\mu_B/T$ values 
0, 1, 2, 3 and 4. It can be seen from the Fig. \ref{fig:chi2B_T_scale} that I-HRG cross the
SB limit in the same temperature regime where PQM show peak like behavior indicating the change
in the relevant degrees of freedom that take part in the thermodynamics of the system. Since I-HRG 
has only hadronic degrees of freedom, it cannot show such non monotonic behavior, exhibiting only 
a monotonic rise with increasing $T$. This lends support to our proposal to estimate the region 
of applicability of the I-HRG model to compute QCD thermodynamics by extracting the boundary where the I-HRG 
model overshoots the corresponding SB value.

\subsection{\label{sec:Result3}Heavy Ion scenario}
Now we will discuss in the context of heavy ion collision. In this scenario $\mu_S$ and $\mu_Q$ 
are non-zeroes and can be calculated applying the following charge conservation equations
\begin{equation}
\label{eq:ns}
\sum_i n_i (T, \mu_B, \mu_S, \mu_Q) S_i=0,
\end{equation}
and
\begin{equation}
\label{eq:nbq}
\frac{\sum_i n_i (T, \mu_B, \mu_S, \mu_Q) Q_i}{\sum_i n_i (T, \mu_B, \mu_S, \mu_Q) B_i} = r,
\end{equation}
where $r$ is the ratio of net-charge to net-baryon number of the colliding nuclei.
For Pb + Pb or Au + Au collisions $r = N_p /(N_p + N_n) \simeq 0.4$, where $N_p$ and $N_n$ are 
respectively proton numbers and neutron numbers of the colliding nuclei.
The right-hand side of the Eq. \ref{eq:ns} is zero since 
initially there is no net-strangeness in the colliding nuclei. 

In Fig. \ref{fig:phase_with_con} we have shown the upper bound for the I-HRG phase on the 
QCD phase diagram for the heavy ion collision scenario. Here charge conservations are applied 
to get $\mu_S$ and $\mu_Q$ at a fixed $T, \mu_B$. Left and right panels show the results using 
the hadron spectra PDG 2016 and PDG 2016+ respectively. It is observed that charge conservation 
does not significantly modify the earlier obtained results at low $\mu_B$ region. However, in 
the intermediate region where both $T$ and $\mu_B$ are large, the imposition of charge conservation 
push the bounds slightly towards higher values of $(\mu_B,T)$.

\subsection{\label{sec:Comparison}Comparison with lattice QCD}

In Fig. \ref{fig:phase_com} we have shown the obtained upper bound for the extent 
of the I-HRG phase in all the four cases studied here. These are the boundaries of 
the shaded regions or the common allowed I-HRG phase by all the studied TQs as shown 
in the previous plots. We have compared our estimates with the hadronic to QGP phase 
transition line (pseudo critical line) calculated by the lattice QCD simulation \cite{Cea:2014xva}. 
The transition line is generally parametrized as
\begin{equation}
 \frac{T(\mu_B)}{T(0)} = 1- \kappa \left(\frac{\mu_B}{T(0)}\right)^2,
\end{equation}
where $T(0)$ and $\kappa$ are the transition temperature and the curvature of the transition
line respectively at zero baryon chemical potential
\cite{Endrodi:2011gv,Kaczmarek:2011zz,Bonati:2014rfa,Cea:2014xva,Bonati:2015bha}. 
The green error band shows the lattice result of transition line
where  $T(0) = 154(9)$ and $\kappa = 0.020(4)$ \cite{Cea:2014xva}.
The yellow box in $T$ at $\mu_{B}$ = 0 reflects the uncertainties associated
with LQCD determination of $T_0$ (145 - 170 MeV) and the choice of observables for
order parameter. Chemical freeze-out parameters of Refs \cite{Cleymans:2005xv} (Cleymans et. al.) 
and \cite{Andronic:2008gu} (Andronic et. al.) are shown by the red and pink colored shaded bands 
respectively. 

At $\mu_B/T=0$, LQCD computations using different combinations of susceptibilities 
of $B$, $Q$ and $S$ upto fourth order demonstrate the breakdown of applicability of I-HRG beyond 
the chiral transition region $154(9)$ MeV~\cite{Bazavov:2013dta}, providing a stronger constraint 
than we have in this study. However, for larger values of $\mu_B/T$ where currently there is no 
LQCD results, our estimates provide a guidance of the region on the QCD phase diagram where one would 
expect the breakdown of the I-HRG model. For $\mu_B/T>2$, $C_{BS}$ provides the strongest bound. 
$C_{BS}$ receives contribution from strange baryons both in the 
numerator as well as in the denominator while strange mesons contribute only to the denominator. 
As we dial up $\mu_B/T$, the contribution from the strange baryons increase while that of the strange 
mesons remain same. This effectively fasten the growth of $C_{BS}$ with $T$ at higher $\mu_B/T$. The 
obtained bound is found progressively at a lower $T$ as compared to the extracted freeze-out curve by 
fitting the I-HRG model to the measured hadron yields~\cite{Cleymans:2005xv,Andronic:2008gu}. This calls 
for revisiting the extraction of the freeze-out curve at these baryon densities with a version of the 
HRG model that goes beyond the ideal limit. It may be noted that HRG with Van der Waals interactions 
can be tuned to yield a $C_{BS}$ that is closer to LQCD results~\cite{Vovchenko:2017zpj}. Hence, it will 
be important to measure $C_{BS}$ at the RHIC BES, FAIR and NICA energies.

\section{\label{sec:Summary}Summary}
One of the primary goals of heavy ion collision experiments is to study
the QCD phase diagram. Differentiating signals of the hadronic and QGP 
phases, smooth crossover transition from a first order transition and those 
of the critical from non-critical region is a challenging problem in the
field. While experimental measurements are difficult and are ongoing,
the QCD based calculations on lattice have numerical challenges at
finite $\mu_{B}$. On the other hand I-HRG model has been very
successful in explaining both the experimental data on yields of
produced hadrons in heavy-ion collisions with a few parameters as well 
as LQCD observables at $\mu_{B}$ = 0 below the chiral transition region. 
The basic idea of the current work was to get an estimate of the region of 
applicability on the QCD phase diagram of the I-HRG model for the study of 
QCD thermodynamics.

For our study we select some TQs that monotonically rise in I-HRG with increasing 
$T$ and $\mu_B$. On the other hand, within QCD these TQs are supposed to attain the 
corresponding SB limiting values as obtained for an ideal gas of massless quarks 
and gluons. As the TQs computed within I-HRG overshoot the SB limit, it indicates 
the importance of going beyond I-HRG, incorporating hadronic interactions and new 
emerging degrees of freedom that are missing in I-HRG and would tame its monotonic 
rise. We obtain the contours where the I-HRG values become equal to the SB limit 
values. This provides an estimate of the region of applicability of the I-HRG model 
to study QCD thermodynamics.

Several bounds on the $(\mu_B, T)$ plane are obtained corresponding to each TQ like 
pressure, energy density, entropy density and susceptibilities of conserved charges.
A hierarchical structure is noted- higher the derivative of the partition function, 
tighter is the bound. From these calculations we have extracted the tightest bound 
below which all the 
studied TQs are smaller than their SB limit values. This then provides an estimate 
of the extent of the I-HRG phase. Further, we have studied the sensitivity of the 
estimates on the systematics due to the input hadron spectrum and different 
treatments of $\mu_S$ and $\mu_Q$ - in one scenario we have assumed 
$\mu_S = \mu_Q =0$ and in the other scenario $\mu_S$ and $\mu_Q$ have been 
calculated applying global charge conservation relevant to the heavy ion collision 
experiment.

At $\mu_B/T=0$, our bound of $T\leq175$ MeV is trivial as current LQCD computations 
already provide a tighter bound of 154(9) MeV~\cite{Bazavov:2013dta}. However, as 
we go to larger $\mu_B/T$ where lattice computations 
plagued by the sign problem become increasingly difficult to perform, our estimates 
provide a 
good starting point. While $\chi^2_Q$ provides the tightest bound for $\mu_B/T<2$, 
$C_{BS}$ takes over for $\mu_B/T>2$. With increasing $\mu_B/T$, the gap between our 
obtained bound and the freeze-out curve extracted from the analysis of the hadron 
yields within the framework of I-HRG widens. Van der Waals type interactions has 
been shown to cure the rise in $C_{BS}$ in I-HRG~\cite{Vovchenko:2017zpj}. This 
probably hints at the importance of such interaction in a baryonic fireball and the 
requirement to analyze the freeze-out curve with such interacting HRG.

\section*{Acknowledgments}

SC acknowledges fruitful discussions with Rohini Godbole, Sourendu Gupta and Harvey B. Meyer.
SC is supported by the AGH UST statutory tasks No.~11.11.220.01/1 within subsidy of 
the Polish Ministry of Science and Higher Education (MNiSW) and the National Science Centre grant 
2015/17/B/ST2/00101. BM acknowledges DST J C Bose fellowship for financial assistance.
SS acknowledges financial support from DAE, Government of India.\\

\section*{References}

\bibliographystyle{h-physrev.bst}
\bibliography{RefFile}

\end{document}